\documentclass[a4paper]{article}

\usepackage{INTERSPEECH2021}
\usepackage{color}
\usepackage{hyperref}

\definecolor{dgreen}{RGB}{0,128,0}
\definecolor{dblue}{RGB}{0,0,128}
\definecolor{dred}{RGB}{128,0,0}
\definecolor{dpurple}{RGB}{128,0,128}

\def\mbf#1{\mathbf{#1}}
\def\mbb#1{\mathbb{#1}}

\def\bs#1{\boldsymbol{#1}}
\def\myx{\mbf{x}}

\def\myz{\mbf{z}}
\def\mys{\mbf{s}}

\newcommand*{\myxc}[1][]{{\color{black}\mathbf{x}_{#1}}}
\newcommand*{\myzc}[1][]{{\color{black}\mathbf{z}_{#1}}}

\usepackage{enumitem}

\title{A Benchmark of Dynamical Variational Autoencoders \\ applied to Speech Spectrogram Modeling}
\name{Xiaoyu Bie$^{1}$, Laurent Girin$^{2}$, Simon Leglaive$^{3}$, Thomas Hueber$^{2}$ and Xavier Alameda-Pineda$^{1}$}
\address{
  $^1$ Inria, Univ.~Grenoble Alpes, CNRS, LJK, 38000 Grenoble, France.\\
  $^2$ Univ.~Grenoble Alpes, CNRS, Grenoble-INP, GIPSA-lab, 38000 Grenoble, France.\\
  $^3$ CentraleSup\'elec, IETR, 35576 Cesson-Sévigné, France.}
\email{}

\begin{document}

\maketitle
\begin{abstract}
  The Variational Autoencoder (VAE) is a powerful deep generative model that is now extensively used to represent high-dimensional complex data via a low-dimensional latent space learned in an unsupervised manner. In the original VAE model, input data vectors are processed independently. In recent years, a series of papers have presented different extensions of the VAE to process sequential data, that not only model the latent space, but also model the temporal dependencies within a sequence of data vectors and corresponding latent vectors, relying on recurrent neural networks. We recently performed a comprehensive review of those models and unified them into a general class called Dynamical Variational Autoencoders (DVAEs). In the present paper, we present the results of an experimental benchmark comparing six of those DVAE models on the speech analysis-resynthesis task, as an illustration of the high potential of DVAEs for speech modeling.
\end{abstract}


\section{Introduction}

The Variational Autoencoder (VAE) introduced in \cite{Kingma2014,rezende2014stochastic} is a powerful deep generative model that is now extensively used to represent high-dimensional data via a low-dimensional latent space learned in an unsupervised manner. It has been used for speech modeling in, e.g., \cite{blaauw2016modeling, hsu2016voice, hsu2017learning, bando2017statistical, leglaive2018variance, akuzawa2018learning, pandey2018monaural, leglaive2019speech, leglaive2019semi}. 

The original VAE does not include temporal modeling. This means that every data vector from a dataset is processed independently of the other data vectors (and the corresponding latent vector is also processed independently of the other latent vectors). In recent years, a series of papers have presented different extensions of the VAE to process sequential data, that not only model the latent space, but also model the temporal dependencies within a sequence of data vectors and corresponding latent vectors, relying on recurrent neural networks (RNNs) \cite{bayer2014learning, fabius2014variational, krishnan2015deep, chung2015recurrent, gu2015neural, fraccaro2016sequential, krishnan2017structured, fraccaro2017disentangled, goyal2017z, hsu2017unsupervised, yingzhen2018disentangled, leglaive2020recurrent}. In practice, those different models vary in how they define the dependencies between the observed and latent variables, how they define and parameterize the corresponding generative distributions, and, importantly, how they define and parameterize the inference model, which is a key ingredient of the VAE methodology. They also differ on how they combine the variables with RNNs to model temporal dependencies, at both generation and inference. In contrast, and remarkably, the training phase is quite similar between models since it is consistently based on the VAE methodology: chaining the inference and generative models (the encoder and decoder) and maximizing a lower bound of the data likelihood over a training dataset. 

In \cite{girin2020dynamical}, we performed an extensive and comprehensive literature review of these models. We introduced a general class of models called Dynamical Variational Autoencoders (DVAEs) that encompasses and unifies the above-cited temporal VAE extensions. The objectives and contributions of the present paper are the following. First, it seems that the DVAE class of models is still relatively poorly known by the speech processing community, yet it has the potential to yield major advances in many speech processing applications such as speech signal synthesis and transformation. So far, it has been used only in a very few studies, see for example the pioneering works in speech coding \cite{yang2020feedback} or speech denoising \cite{leglaive2020recurrent}. 
Therefore we want to disseminate the existence of the DVAE class of models to the speech processing community and foster research addressing speech processing applications with DVAEs. Then, as an illustration of their potential for speech modeling, in the present paper, we report the results of an experimental benchmark conducted on the speech (power spectrogram) analysis-resynthesis task. We selected six of the DVAE models that we detailed in~\cite{girin2020dynamical}, we reimplemented them and compared their performance on this task. The PyTorch code is made publicly available, and we believe it can be of interest for many researchers interested in joint unsupervised representation learning and dynamical modeling of speech signals.


\section{Dynamical VAEs}

\subsection{The original VAE model}

The seminal VAE model introduced in \cite{Kingma2014,rezende2014stochastic} is defined by:
\begin{equation}
p_{\bs{\theta}}(\mbf{x}, \mbf{z}) = p_{\bs{\theta}}(\mbf{x} | \mbf{z})p(\mbf{z}),
\label{joint_dist_y_z}
\end{equation}
where $p(\mbf{z})$, the prior distribution of the latent variable $\mbf{z}$, is a multivariate standard Gaussian distribution, $p_{\bs{\theta}}(\mbf{x} | \mbf{z})$ is the (conditional) \emph{likelihood function} of the observed variable $\mbf{x}$, and the dimension $L$ of $\mbf{z}$ is (much) lower than the dimension $F$ of $\mbf{x}$. The parameters of $p_{\bs{\theta}}(\mbf{x} | \mbf{z})$ are provided by a deep neural network (DNN), called the decoder network, that takes $\mbf{z}$ as input. $\bs{\theta}$ represents the parameters of this decoder network (e.g., the weights and biases of a multi-layer perceptron). 


Because the relationship between $\mbf{z}$ and $\mbf{x}$ is highly non-linear, the posterior distribution $p_{\bs{\theta}}(\mbf{z} | \mbf{x})$ is not analytically tractable. It is thus approximated with a parametric variational distribution $q_{\bs{\phi}}(\mbf{z} | \mbf{x})$, a.k.a.\ the inference model, whose parameters are provided by another DNN (called the encoder network, with weights $\bs{\phi}$ and input $\mbf{x}$). 
A usual choice is to set $q_{\bs{\phi}}(\mbf{z} | \mbf{x})$ as a Gaussian distribution with diagonal covariance matrix. 
The parameters $\{\bs{\theta},\bs{\phi}\}$ are then jointly estimated by  maximizing a lower bound of the data log-likelihood function, called the Variational Lower Bound (VLB) or Evidence Lower Bound (ELBO), given by (for one single data vector):
\begin{align}
\mathcal{L}(\bs{\phi}, \bs{\theta}, \mbf{x}) &=  \mathbb{E}_{q_{\bs{\phi}}(\mbf{z} | \mbf{x})}\big[ \ln p_{\bs{\theta}}(\mbf{x} | \mbf{z}) \big] \! - \! D_{\text{KL}}\big[q_{\bs{\phi}}(\mbf{z} | \mbf{
x}) \parallel p(\mbf{z})\big], \label{eq:VAE-VLB-a}
\end{align} 
and evaluated on a training dataset ($D_{\text{KL}}$ denotes the Kullback-Leibler divergence). Maximization of the VLB is done by combining stochastic gradient descent with sampling techniques. Such optimization process is now considered as routine within deep learning toolkits such as Keras and PyTorch.

\subsection{From VAE to DVAEs}
\label{subsec:VAE2DVAE}

As already mentioned in the introduction, the Dynamical Variational Autoencoder (DVAE) is a class of deep generative models that generalizes the VAE to the modeling of sequential data \cite{girin2020dynamical}. 
The DVAE models that we consider here process an ordered time sequence of vector data $\mbf{x}_{1:T}= \{\mbf{x}_{t}\}_{t=1}^T$ and a corresponding ordered sequence of latent vectors $\mbf{z}_{1:T} = \{\mbf{z}_{t}\}_{t=1}^T$. 
A DVAE is thus defined by the following joint probability density function (pdf), which is  a generalization of \eqref{joint_dist_y_z}:
\begin{align}
p_{\bs{\theta}}(\mbf{x}_{1:T}, \mbf{z}_{1:T}) &= p_{\bs{\theta}}(\mbf{x}_{1:T} |  \mbf{z}_{1:T})p_{\bs{\theta}}(\mbf{z}_{1:T}).
\label{joint_dist_y_z_DVAE}
\end{align}
However, this form does not give much information about the generative process, and we prefer to use the chain rule to reformulate \eqref{joint_dist_y_z_DVAE} as:
\begin{align}
p_{\bs{\theta}}(\mbf{x}_{1:T}, \mbf{z}_{1:T}) &= \prod_{t=1}^{T} p_{\bs{\theta}}(\mbf{x}_{t} | \mbf{x}_{1:t-1},\mbf{z}_{1:t}) p_{\bs{\theta}}(\mbf{z}_{t} | \mbf{x}_{1:t-1},\mbf{z}_{1:t-1}).
\label{time_slide_ordered_model}
\end{align}
This particular reformulation (among many other possibilities) is a \emph{causal} form: $\mbf{z}_{t}$ is first generated from $\mbf{z}_{1:t-1}$ and $\mbf{x}_{1:t-1}$, and then $\mbf{x}_{t}$ is generated from $\mbf{x}_{1:t-1}$ and $\mbf{z}_{1:t}$. In a general manner, the dependencies are implemented using RNNs~: The parameters of the generative distributions of  $\mbf{z}_{t}$ and $\mbf{x}_{t}$ are the outputs of RNNs that take $\mbf{z}_{1:t-1}$ and $\mbf{x}_{1:t-1}$ (and $\mbf{z}_{t}$ for the generation of $\mbf{x}_{t}$) as inputs.


The different DVAE models that we cited in the introduction are all special cases of the general expression \eqref{time_slide_ordered_model} where the dependencies are possibly simplified. In the following of the paper, we consider the six following DVAE models: The Deep Kalman Filter model (DKF) \cite{krishnan2015deep,krishnan2017structured}, the Stochastic Recurrent Neural Network (STORN) \cite{bayer2014learning}, the Variational Recurrent Neural Network (VRNN) \cite{chung2015recurrent, goyal2017z}, another type of Stochastic Recurrent Neural Network (SRNN) \cite{fraccaro2016sequential}, the Recurrent Variational Autoencoder (RVAE) \cite{leglaive2020recurrent} and the Disentangled Sequential Autoencoder (DSAE) \cite{yingzhen2018disentangled}. The corresponding simplified forms of the generative distributions are given in Table~\ref{tab:DVAE-pdfs}. Note that VRNN is the ``richer'' possible DVAE model in terms of variable dependencies since all dependencies in \eqref{time_slide_ordered_model} are kept, whereas in contrast, the original VAE can be seen as a DVAE where all temporal dependencies have been removed.

\begin{table}[th]
\caption{Conditional independence assumptions for various models in the DVAE family. The * indicates that the inference model is compliant with the structure of the true posterior. For DSAE, $\mbf{v}$ is an additional sequence-level variable (not detailed here, see \cite{yingzhen2018disentangled, girin2020dynamical} for details).}
\label{tab:DVAE-pdfs}
\centering
\resizebox{\linewidth}{!}{
\begin{tabular}{llll}
\toprule
& & $p_{\theta}(\myzc[t] | \myxc[1:t-1],\myzc[1:t-1])$ & $p_{\theta}(\myxc[t] | \myxc[1:t-1],\myzc[1:t])$ \\[.1cm]
\midrule
VAE\textsuperscript{*} & \cite{Kingma2014,rezende2014stochastic}  & $p_{\boldsymbol\theta}(\myzc[t])$ & $p_{\boldsymbol\theta}(\myxc[t] | \myzc[t])$ \\
RVAE\textsuperscript{*} & \cite{leglaive2020recurrent} & $p_{\boldsymbol\theta}(\myzc[t])$ & $p_{\boldsymbol\theta}(\myxc[t] | \myzc[1:t])$ \\
STORN & \cite{bayer2014learning} & $p_{\boldsymbol\theta}(\myzc[t])$ & $p_{\boldsymbol\theta}(\myxc[t] | \myxc[1:t-1], \myzc[1:t])$\\
DKF\textsuperscript{*} & \cite{krishnan2015deep,krishnan2017structured} & $ p_{\boldsymbol\theta}(\myzc[t] | \myzc[t-1]) $ & $p_{\boldsymbol\theta}(\myxc[t] | \myzc[t])$\\
DSAE & \cite{yingzhen2018disentangled} & $ p_{\boldsymbol\theta}(\myzc[t] | \myzc[1:t-1]) $ & $p_{\boldsymbol\theta}(\myxc[t] | \myzc[t], \mbf{v})$\\
VRNN & \cite{chung2015recurrent, goyal2017z} & $ p_{\boldsymbol\theta}(\myzc[t] | \myxc[1:t-1], \myzc[1:t-1]) $  & $p_{\boldsymbol\theta}(\myxc[t] | \myxc[1:t-1], \myzc[1:t])$\\
SRNN\textsuperscript{*} & \cite{fraccaro2016sequential} & $ p_{\boldsymbol\theta}(\myzc[t] | \myxc[1:t-1], \myzc[1:t-1]) $ & $p_{\boldsymbol\theta}(\myxc[t] | \myxc[1:t-1], \myzc[t])$\\[.1cm]
\bottomrule
\end{tabular}
}
\end{table}

The inference and training methodology of a DVAE model follows the one of the VAE: Definition of an inference model $q_{\bs{\phi}}(\mbf{z}_{1:T} | \mbf{x}_{1:T})$ (since the exact posterior distribution $p_{\bs{\theta}}(\mbf{z}_{1:T} | \mbf{x}_{1:T})$ is not analytically tractable), chaining of the encoder and decoder, and training by maximizing the VLB on training data. Similar to the generative model, it is convenient to reshape the inference model in the following general form, using the chain rule:
\begin{align}
q_{\bs{\phi}}(\mbf{z}_{1:T} | \mbf{x}_{1:T}) &= \prod_{t=1}^{T} q_{\bs{\phi}}(\mbf{z}_{t} | \mbf{z}_{1:t-1},\mbf{x}_{1:T})
\label{DVAE_inference_model_time_ordered}.
\end{align}
We can see that \eqref{DVAE_inference_model_time_ordered} is causal  regarding the past latent vectors $\mbf{z}_{1:t-1}$ but not regarding the complete sequence of observed vectors $\mbf{x}_{1:T}$. Similar to the DVAE generative model, the dependencies in \eqref{DVAE_inference_model_time_ordered} can be simplified (or not), depending on, e.g., if one wants the inference model to have the same structure as the exact posterior distribution \cite{geiger1990identifying}, \cite[Chapter 8]{BishopBook}, or if one wants to have a causal implementation to enable online inference. 
In the present paper, for each DVAE model we used the inference model defined in the corresponding original paper.

Still Similar to the VAE, the training of DVAE models is based on maximization of the VLB, here defined by (for one data sequence) \cite{girin2020dynamical}:
\begin{align}
\mathcal{L}(\bs{\phi}, \bs{\theta}, \mbf{x}_{1:T}) &= \mbb{E}_{q_{\bs{\phi}}(\mbf{z}_{1:T} | \mbf{x}_{1:T})} \big[ \ln p_{\bs{\theta}}(\mbf{x}_{1:T},\mathbf{z}_{1:T}) \nonumber 
\\ & \quad - \ln q_{\bs{\phi}}(\mathbf{z}_{1:T} | \mbf{x}_{1:T}) \big].
\label{VFE-DVAE-general}
\end{align}
The developed form of this VLB, obtained by reinjecting \eqref{time_slide_ordered_model} and \eqref{DVAE_inference_model_time_ordered} into \eqref{VFE-DVAE-general} and using some ``cascade'' trick, is given in \cite{girin2020dynamical}. Of course, depending on the specific (chosen) DVAE generative and inference models, this developed form can be simplified. In a general manner, expressing the VLB in a form that is differentiable w.r.t.~$\bs{\phi}$ and $\bs{\theta}$ requires some sampling of $\mbf{z}_{1:T}$, which is here done recursively. This sampling is alternated with calculation of the VLB gradient over a training dataset and parameter update. Again, see \cite{girin2020dynamical} for more detailed information. 


\section{Application to speech power spectrogram modeling}
\label{sec:VAE_speech_models}


In the literature, the DVAE models have been applied to different kinds of data. Here we focus on the modeling of speech signals. This is done in the short-term Fourier transform (STFT) domain: The time-domain speech waveform is first transformed into a speech STFT spectrogram  $\mbf{s}_{1:T} = \{\mbf{s}_{t}\}_{t=1}^T$, where each complex-valued vector $\mbf{s}_{t} = \{s_{t,f}\}_{f=0}^{F-1}$ is the speech short-term spectrum at time index $t$, and $f$ is the frequency bin. As is usual in speech/audio processing, $s_{t,f}$ is assumed to follow a circular-symmetric zero-mean complex-valued Gaussian distribution, see, e.g., \cite{ephraim1984speech, fevotte09, liutkus2011gaussian}. Moreover, the STFT coefficients at different frequency bins are assumed to be (conditionally) independent, i.e.~the covariance matrix of $\mbf{s}_{t}$ is diagonal. 

In practice, the data sequence $\myx_{1:T}$ processed by a DVAE is the STFT \textit{power} spectrogram, i.e.~$x_{t,f} = |s_{t,f}|^2$ for all time-frequency bins. Given the above assumption on $s_{t,f}$, each power spectrogram coefficient $x_{t,f}$ follows a Gamma distribution with shape parameter $1$ and scale parameter ${\sigma}_{\mys,t,f}^2(\mbf{x}_{1:t-1},\mbf{z}_{1:t})$
\cite{fevotte09, fevotte09eusipco, girin2019notes}.\footnote{$\sigma_{\mys,t,f}^2(\cdot)$ is also the variance of $s_{t,f}$, the mean of $x_{t,f}$, and the speech signal power spectral density.} In the most general DVAE context, the parameter vector $\bs{\sigma}_{\mbf{s},t}^2(\cdot) = \{\sigma_{\mys,t,f}^2(\cdot)\}_{f=0}^{F-1}$ depends on $\mbf{x}_{1:t-1}$ and $\mbf{z}_{1:t}$, i.e.~it is provided by RNNs taking $\mbf{x}_{1:t-1}$ and $\mbf{z}_{1:t}$ as inputs. Again, those dependencies can be simplified depending on the specific DVAE model, see Table~\ref{tab:DVAE-pdfs}.  
Note that by using the above probabilistic model, we  assume that the phase of $s_{t,f}$ follows a uniform distribution in $[0, 2\pi[$.
This is a very common assumption in speech/audio processing, since modeling the phase of STFT spectrograms is still a very challenging task, be it with classical statistical models or with deep-learning-based models (see \cite{nugraha2019deep} for an example in the VAE framework).

As for the latent vector generative distribution $p_{\bs{\theta}}(\mbf{z}_{t} | \mbf{x}_{1:t-1},\mbf{z}_{1:t-1})$, it is set in its most general form as a (real-valued) Gaussian distribution, with mean vector $\bs{\mu}_{\mbf{z},t}(\mbf{x}_{1:t-1},\mbf{z}_{1:t-1})$ and a diagonal covariance matrix with entries from vector $\bs{\sigma}_{\mbf{z},t}^2(\mbf{x}_{1:t-1},\mbf{z}_{1:t-1})$. Those two vectors are provided by RNNs taking here $\mbf{x}_{1:t-1}$ and $\mbf{z}_{1:t-1}$ as inputs, and again, the dependencies can be simplified according to Table~\ref{tab:DVAE-pdfs} depending on the specific DVAE model.
Note that in all cases the entries of the latent vector $\mbf{z}_t$ are assumed (conditionally) independent, which is in line with the principle of looking for a disentangled latent representation \cite{girin2020dynamical}. 

Finally, the encoder follows \eqref{DVAE_inference_model_time_ordered}, where, in a general manner, $q_{\bs{\phi}}(\mbf{z}_{t} | \mbf{z}_{1:t-1}, \mbf{s}_{1:T})$ is a (real-valued) Gaussian distribution with mean vector $\bs{\mu}_{\bs{\phi}}(\mbf{z}_{1:t-1}, \mbf{s}_{1:T})$ and a diagonal covariance matrix with entries from vector $\bs{\sigma}_{\bs{\phi}}^2(\mbf{z}_{1:t-1}, \mbf{s}_{1:T})$. 
Those two vectors are provided by the encoder RNN. As stated in Section~\ref{subsec:VAE2DVAE}, the dependencies can be simplified, and in our experiments, for each specific DVAE model, we used the inference model described in the corresponding original paper.

\section{Experimental benchmark}
\label{sec:experiments}

In the following, we present our experimental benchmark of the six DVAE models of Table~\ref{tab:DVAE-pdfs} applied to speech power spectrogram analysis-resynthesis. In short, a speech power spectrogram $\myx_{1:T}$, is encoded into, and then resynthesized from, a latent vector sequence $\myz_{1:T}$. 

\subsection{Implementation of the DVAE models}
\label{subsec:model-implementation}

Because of lack of room, it is not possible to detail here the implementation of each model. The detailed implementation equations of both the generative part and inference part of each DVAE model, involving  the expression of RNN internal states, are detailed in \cite{girin2020dynamical}.  Importantly, not only the different DVAE models  differ in the way the dependencies in $p_{\bs{\theta}}(\mbf{x}_{t} | \mbf{x}_{1:t-1},\mbf{z}_{1:t})$ and $p_{\bs{\theta}}(\mbf{z}_{t} | \mbf{x}_{1:t-1},\mbf{z}_{1:t-1})$ are simplified, but a given DVAE model can have different implementations. Indeed, the parameters of the generative distributions are provided by neural networks, and many different network implementations can be considered for the same dependency structure  \cite{girin2020dynamical}. In a general manner, we have tried to find a good trade-off between respecting the architecture of the model as described in the original paper and ensuring a fair comparison between the different models for the speech analysis-resynthesis task. Similar modules across different DVAE models are thus implemented in the same manner, i.e.~with the same number of layers, units per layers, and activation function (see \cite{girin2020dynamical} for details). 
Moreover, the following specifications are common to all DVAEs:
\begin{itemize}[leftmargin=*]
    \item The dimension of the observation vector $\myx_t$ and output parameter $\bs{\sigma}_{\mbf{s},t}^2(\cdot)$ is set to $F = 257$ (see next subsection);
    \item The dimension of the latent vector $\myz_t$ is set to $L = 16$;
    \item The dimension of RNN hidden internal state vectors is set to $128$; Unless specified in the original paper, all RNNs are instantiated as LSTM networks;
\end{itemize}


\subsection{Dataset and pre/post-processing} 

We used the Wall Street Journal (WSJ0) dataset \cite{WSJ0}.
The \textit{si\_tr\_s}, \textit{si\_dt\_05} and \textit{si\_et\_05} subsets were used for model training, validation, and test, respectively. The STFT was applied on 16-kHz signals with a $32$-ms sine window ($512$ samples) and $50$\%-overlap to obtain sequences of $257$-dimensional discrete spectra (for positive frequencies). We set $T=150$ (i.e.~$2.4$-s speech sequences). In summary, each data sequence is a $150 \times 257$ STFT power spectrogram. This data preprocessing resulted in a set of $N_{\rm{tr}} = 13,272$ training sequences (about $9$h of speech signal) and $N_{\rm{val}} = 2,143$ validation sequences (about $1.5$h). For test, we used the STFT power spectrogram of each complete test sequence (with beginning and ending silence portions removed), which can be of variable length, most often larger than $2.4$s (total of about $1.5$h). 
For the evaluation, we used the reconstructed power spectrograms, as well as the reconstructed waveforms, obtained by combining the reconstructed magnitude spectrograms with the input phase spectrograms and applying inverse STFT with overlap-add.

\subsection{Training and testing}
 
All models were implemented in PyTorch~\cite{paszke2019pytorch}. Training was made with mini-batch stochastic gradient descent, using the Adam optimizer~\cite{Adam}, with a learning rate of 0.0001 and a batch size of $32$. We used early stopping on the validation set with a patience of 20 epochs. After training, we evaluated the average performance on the test set using the following three metrics: The root mean squared error (RMSE) between original and reconstructed waveforms,\footnote{Because of the orthogonal properties of the Discrete Fourier Transform, and because the phase of the original spectrogram is combined with the reconstructed magnitude spectrogram, RMSE calculated between speech waveforms is equivalent to RMSE calculated between the corresponding magnitude spectrograms.} Perceptual Evaluation of Speech Quality (PESQ) scores \cite{rix2001perceptual} and Short-Time Objective Intelligibility (STOI) scores \cite{taal2010short}. The amplitude of each original speech waveform was normalized in $[-1, 1]$, so the RMSE (generally much lower than $1$) directly represents a percentage of the maximum amplitude value. PESQ scores are in $[-0.5, 4.5]$ and STOI scores are in $[0, 1]$. For both, the higher the better.

\subsection{Results and discussion}
\label{sec:results}


We first checked that the loss curves (i.e., VLB up to a constant term) obtained on the training data and the validation data show a successful convergence of the training for all the implemented DVAE models. 
Then we report in Table~\ref{tab:comparison} the values of the three evaluation metrics averaged over the test dataset. From this table, and from the observation of reconstructed spectrograms (not shown here because of room limitation), we can draw the following comments:
\begin{itemize}[leftmargin=*]
\setlength\itemsep{0em}
    \item First, all tested DVAE models lead to correct signal reconstruction, with an RMSE that represents only a few percents (generally lower than 5\%) of the maximum waveform amplitude. The quality of the reconstructed speech signals, as measured by PESQ, goes from fair to good. STOI scores, generally higher than $0.90$ show their good intelligibility.
    Importantly, all DVAE models outperform the standard VAE model. This demonstrates the interest of including temporal modeling in the VAE framework for modeling speech signals.
    \item VRNN and SRNN are the two methods with highest performance, with a notable gain in performance over all other models, and SRNN is slightly better than VRNN. We recall that VRNN keeps all possible dependencies in the general DVAE formulation \eqref{time_slide_ordered_model}, and Table~\ref{tab:DVAE-pdfs} shows that SRNN contains more dependencies with the past observed and latent variables than the other implemented DVAEs. We believe that this allows VRNN and SRNN to better capture the temporal patterns of speech spectrograms.
As for SRNN performing slightly better than VRNN, this could be due to the fact that the inference model of SRNN respects the structure of the exact posterior distribution, which depends on all observations $\myx_{1:T}$, whereas the inference model of VRNN (as proposed in the original paper) does not: it uses only the causal observations $\myx_{1:t}$ \cite{chung2015recurrent, fraccaro2016sequential, girin2020dynamical}.
    \item The performance scores of DKF and STORN are quite equivalent, but below those of SRNN and VRNN. This is likely due to the fact that the temporal dependencies in DKF and STORN are less rich than in VRNN and SRNN. DKF has the structure of a state-space model, where there is no explicit temporal dependency between $\myx_{t-1}$ and $\myx_t$, but only between $\myz_{t-1}$ and $\myz_t$. In STORN $\myx_t$ depends on $\myx_{1,t-1}$ and $\myz_{1:t}$, so one could think that this model is richer that DKF and should have better performance. However, we found out that DKF slightly outperforms STORN in terms of PESQ and STOI (but not in RMSE). We can hypothesize that, here also the difference in performance is (at least partly) due to the fact that the inference model of DKF does respect the structure of the exact posterior distribution whereas the inference model of STORN does not (for STORN, the inference of $\myz_t$ depends only on $\myx_{1:t}$ and not on $\myz_{t-1}$ nor $\myx_{t+1:T}$, see \cite{bayer2014learning, girin2020dynamical}).  Another possible explanation is that the prior distribution of $\myz_t$ is i.i.d.\ in STORN, while it has temporal dependencies in DKF. In a general manner, we hypothesize that models with i.i.d.\ prior over time on $\myz_t$ risk to underperform w.r.t.\ models that are defined via a temporal generative model of $\myz_t$. We must keep in mind that the $\myz_{1:T}$ sequence is assumed to encode high-level characteristics of the data $\myx_{1:T}$ that generally evolve smoothly over time (at least for some of these characteristics). This is not ensured by the i.i.d.~standard Gaussian prior distribution of $\myz_t$ used in STORN. 
    \item The performance of DSAE is quite disappointing, especially compared to the performance of DKF. Indeed, like DKF, DSAE also has the structure of a state-space model. Actually, DSAE can be seen as an improved version of DKF, with an additional sequence-level variable (not detailed here) and infinite-order temporal dependency of $\myz_t$ (as opposed to first-order for DKF). Again, this poor performance could come from the structure of the inference model, which depends on $\myx_{1:T}$, whereas the exact posterior distribution of $\myz_t$ depends on $\myz_{1:t-1}$ and $\myx_{t:T}$ \cite{girin2020dynamical}. 
    
    
    \item In our experiments, RVAE exhibits the worst performance of all tested DVAE models. Here also, i.i.d.~modeling of $\myz_t$ may be suboptimal. In addition to that, there is no explicit modeling of the temporal dependencies on $\myx_t$ (e.g., $\myx_t$ does not depend on $\myx_{t-1}$), hence leading to a model with weak ``predictive power.'' However, we recall that for the present experiments, we set up the neural network architectures so that all compared DVAE models have a similar number of parameters. For RVAE, the architecture was made more complex than in the original paper \cite{leglaive2020recurrent}, which drastically decreased the performance. With the original RVAE model in \cite{leglaive2020recurrent}, the RMSE, PESQ and STOI performance are 0.0297, 3.47 and 0.95, respectively. This shows that two DVAE models with the same probabilistic dependencies but different neural network architectures can perform very differently.
    
\end{itemize}


\begin{table} [t]
\caption{Performance of the tested DVAE models in our speech analysis-resynthesis experiments (RMSE, PESQ and STOI scores averaged over the WSJ0 test subset).}
\label{tab:comparison}
\centering
\resizebox{\columnwidth}{!}{
\begin{tabular}{l c c c c c c c}
\toprule 
& VAE & DKF & STORN & VRNN & SRNN & RVAE & DSAE \\
\midrule
RMSE ($\times 10^{-2}$) & 5.10 & 3.44 & 3.38 & 2.67 & \bf{2.48} & 4.99 & 4.69 \\
PESQ & 2.05 & 3.30 & 3.05 & 3.60 & \bf{3.64} & 2.27 & 2.32 \\
STOI & 0.86 & 0.94 & 0.93 & 0.96 & \bf{0.97} & 0.89 & 0.90 \\
\bottomrule
\end{tabular}
}
\end{table}

\section{Conclusions}

We presented a benchmark of several DVAE models that are of high interest for speech modeling. This benchmark shows that, in a practical application requiring the modeling of speech power/magnitude spectrograms, VRNN or SRNN seem particularly promising. Some of the other tested DVAE models show a slightly lower performance but they also show a reduced complexity that can be an interesting feature. Also, we can conjecture that having an inference model that respects the exact variable dependencies at inference time is very important for obtaining optimal performance. However this is not always possible, e.g.~some applications may require a causal inference model for online processing.

We insist on the fact that the above ``model ranking'' is valid only for the presented experiments, which consist of pure analysis-resynthesis of speech power spectrograms. For other tasks such as speech signal generation (from new values of the latent vectors) or speech signal transformation (with modification of the latent vectors), we have no claim on how the implemented models would behave. So far, it is still quite difficult to know how much of the information about $\myx_{1:T}$, and what kind of information, is encoded into $\myz_{1:T}$, and in particular, we do not know about the 
``disentanglement power'' of each model. And importantly, there is no clear methodology to address those issues, i.e., to evaluate the generated data and the representation power of the latent variable. This is part of our current works.

The code re-implementing the six tested DVAE models (+ the VAE) and used on the benchmark task is made available to the community. The open-source code and the best trained models can be downloaded at the following repository: \url{https://github.com/XiaoyuBIE1994/DVAE-speech}. We have taken care, in the code, to follow the unified presentation and notation used in our review paper \cite{girin2020dynamical}, making it, hopefully, a useful resource to the speech processing community. 

\section{Acknowledgements}

This work has been partially supported by MIAI@Grenoble Alpes (ANR-19-P3IA-0003) and by the European Commission (H2020 SPRING project under GA \#871245).

\bibliographystyle{IEEEtran}

\bibliography{DVAE.bib}


\end{document}